\documentclass[english]{article}
\usepackage[T1]{fontenc}
\usepackage[latin9]{inputenc}
\usepackage[letterpaper]{geometry}
\usepackage{amsmath}
\usepackage{amssymb}

\makeatletter
\newcommand{\lyxaddress}[1]{
\par {\raggedright #1
\vspace{1.4em}
\noindent\par}
}

\makeatother

\makeatother

\usepackage{babel}

\begin{document}

\title{\textbf{Criteria to observe mesoscopic emergence of protein biophysical
properties.}}

\author{\textbf{Anirban Banerji}$^{\text{1}}$, \textbf{Indira Ghosh$^{\text{2*}}$}}

\maketitle

\lyxaddress{\textbf{$^{\text{[1]}}$Bioinformatics Centre, University of Pune},
\textbf{Pune-411007, Maharashtra, India. }\textbf{\underbar{(anirbanab@gmail.com)}}}

\lyxaddress{\textbf{$^{\text{[2]}}$School of Information Technology, Jawaharlal
Nehru University},\textbf{ New Delhi-110067, India.} \textbf{\underbar{(indirag@mail.jnu.ac.in)}}}
\begin{abstract}
Proteins are regularly described with some general indices (mass fractal
dimension, surface fractal dimension, entropy, enthalpy, free energies,
hydrophobicity, denaturation temperature etc..), which are inherently
statistical in nature. These general indices emerge from innumerable
(innately context-dependent and time-dependent) interactions between
various atoms of a protein. Many a studies have been performed on
the nature of these interatomic interactions and the change of profile
of atomic fluctuations that they cause. However, we still do not know,
under a given context, for a given duration of time, how does a macroscopic
biophysical property emerge from the cumulative interatomic interactions.
An exact answer to that question will involve bridging the gap between
nano-scale distinguishable atomic description and macroscopic indistinguishable
(statistical) measures, along the mesoscopic scale of observation.
In this work we propose a computationally implementable mathematical
model that derives expressions for observability of emergence of a
macroscopic biophysical property from a set of interacting (fluctuating)
atoms. Since most of the aforementioned interactions are non-linear
in nature; observability criteria are derived for both linear and
the non-linear descriptions of protein interior. The study assumes
paramount importance in $21^{st}$-century biology, from both the
theoretical and practical utilitarian point of view. While it helps
the theoretical discourse by providing a framework to understand the
origin of a macroscopic property; ability of it to predict $a\, priory$
whether the dynamics in a certain set of atoms or the couplings between
them, can at all produce a biological property of interest or not,
will account for tremendous saving of resource and effort.\\
 \\
 \\
 \\
 \\
 \\

\end{abstract}

\section{\underbar{Introduction :}}

Recent works have described proteins as 'complex systems'{[}1, 2{]}
and as 'deformable polymers'{[}3{]}. The mesoscopic nature of protein
structures has been reported by crystallographers too {[}4{]}. Furthermore,
it has been found recently that proteins exist in a state of 'self
organized criticality'{[}5, 6{]}. Along with all these, recent {[}7{]}
and previous {[}8{]} characterizations of inhomogeneous distributions
of mass and hydrophobicity merely serve to complicate an effort to
construct a general and unambiguous scheme for description of protein
interior. An approach to study protein interior that describes the
inhomogeneous, nonlinear behaviors of protein structural parameters
can be constructed by describing them through self similarity. Indeed
many previous studies on this topic (a dreadfully undersized representation
is references {[}9{]}, {[}10-18{]}) had hinted that with an objective
quantification of self-similarity, we can decipher the hidden symmetry,
which connects global patterns of macroscopic properties in proteins
(say hydrophobicity distribution, polarizability distribution etc..)
with the local (atomic) interactions that produce them {[}19{]}.\\
 \\
 However the basic question, that, precisely when does the macroscopically
measurable quantities emerge from the microscopic interactions between
the atoms, - could not be answered from any of these approaches. Such
an examination of protein interior is necessary not only for purely
theoretical discourse, but also for emerging practical applications
that attempt to describe proteins from paradigms of nanotechnology
and nano-science and mesoscopic-science. Protein function is a dynamic
property that comes into being due to conformational changes of protein
structure in its physiological environment. Hence, as have been commented
upon in a recent study{[}20{]}, to understand and control the function
of target proteins, it assumes paramount importance to develop methods
that can analyze the collective motions at the molecular level, from
which the macroscopic properties emerge. This question is therefore
of immense importance to contemporary protein designing and protein
engineering studies. The aforementioned question can be alternatively
posed from the perspective of control systems study; viz. when does
a particular biophysical property become observable in a sub-set of
protein atoms? Re-framed equivalently : whether we can call a sub-system
of protein atoms to be observable with respect to a particular biophysical
property or not? - This precise question can obviously be answered
in three ways; viz. the sub-system of atoms under question, is completely
observable with respect to the biophysical property under consideration;
it is partially observable, or it is not observable at all. Derivations
of the precise mathematical conditions for these three cases, form
the focus of the present work. Algorithmic implementations of these
mathematical conditions can easily be achieved to relate the general
theoretical framework to particular simulation oriented studies that
can be applied directly to the practical situations.\\
 \\
 Since a protein is comprised of large number of atoms, a realistic
scheme that attempts to describe any global biophysical property (say
free energy of a protein, radius of gyration of it, hydrophobic fractal
dimension of it, resultant dipole moment of it, etc ..) from the study
of relevant property associated with individual atoms (say, individual
spatial fluctuation of each atom, mass and volume of an atom, partial
charge in each of the atoms, etc ..) becomes difficult to construct,
analyze and understand. The need therefore, is to construct a method
that is mathematically accurate, but at the same time, easily implementable
algorithmically. A method that can reduce the shear scale of dimensionality
(huge number of atoms, many properties, etc ..) associated with the
problem. Such a method can only be realized, if it can detect commonalities
in patterns across different biophysical properties. Since the properties
under consideration are all macroscopic in nature, the problem becomes
especially difficult to pose if studies with lower number of atoms
than the permitted threshold for emergence of the property are attempted.
However, although complicated, construction of this algorithm holds
enormous importance for several paradigms of protein biophysics. A
general mathematical construct to address this problem should be able
provide a simple yet reliable framework to describe and analyze the
connection between individual properties at the atomic scale and globally
emergent protein scale. In this work, we propose such an algorithm.
Indeed over the years some attempts have been made in the paradigm
of protein biophysics, to establish the relationship between coupled
microscopic fluctuations and their effect on causing the macroscopic
behaviour {[}21, 22{]}; but scope of these efforts were limited to
certain specialized fields and were not general. The present work,
on the other hand attempts to construct the generalized conditions
to achieve the same.\\
 \\
 Since the number of works that have attempted to view proteins
as mesoscopic systems are less, it assumes importance at this point
to clarify the exact goal of the present paper. Macroscopic states
describe a system from a top-down perspective with variables of (mostly)
statistical origin. In other words, macroscopic description of the
system contains at lower level of information than the same with microscopic
description. However, since the measurable variables themselves are
macroscopic, a macroscopic description still meets the given requirements
of accuracy. In the realm of protein biophysics, the consideration
of macroscopic states like temperature, pressure, enthalpy or entropy
is sufficient to describe the behavior of the system and knowledge
of microscopic states like position and velocity of each of the involved
molecules is not always necessary. The transition from microscopic
level of description to the macroscopic one, however, is not continuous
and if we merely consider these two modes of description of a system
(microscopic and macroscopic), the transition takes place in a step-function
like discontinuous mode at the statistical parametric limit (number
of components of the system = 32). - Such a scheme of description
of biophysical properties might not always be correct. Properties
do not emerge suddenly, but appropriates special features and characteristics
gradually as the description turns gradually from microscopic to macroscopic
paradigm. Mesoscopic states are states containing the intermediate
details. It is at this particular scale that we can expect to observe
the origin and gradual coming to being of (most of the) biophysical
properties. Hence, it assumes enormous importance to construct objective
frameworks compatible to the mesoscopic scale of protein description,
so that one would be able to scrutinize the multifaceted characteristics
of the origin and development of the biophysical property of his/her
interest.\\
 \\
 The basic and only assumption of the present work is that the
emergence of macroscopic properties can be studied with accuracy and
consistency, by studying the interatomic interaction profiles in sufficient
details. Assertions from some recent works {[}23-25{]} support this
assumption strongly. Interatomic interactions manifest themselves
through their fluctuation profiles. Interatomic fluctuations are so
important to protein's existence that any instantaneous conformation
of its, is known to fluctuate thermally around its native conformation.
In the interior of the protein, atoms are tightly packed and the interactions
between the atoms assume complicated nature, but the fluctuations
prevail there too. These fluctuations have been studied from various
perspectives by various works. The thermal, conformational fluctuations
of a globular protein were decomposed into collective motions and
studied from various perspectives {[}26-30{]}. In normal modes analysis,
the fluctuations are expressed by a linear combination of normal modes
{[}28-30{]}. However, construction of a method to observe the emergence
of a macroscopically measured property from the microscopic fluctuations
through a mesoscopic limit, had never been tried before. The present
algorithm attempts to trace back any macroscopically measured property
of statistical origin to time-dependent and context-dependent microscopic
fluctuations, to observe at which mesoscopic limit of number of atoms
does the property emerge and how it grows gradually, before attaining
its macroscopic statistical nature.\\
 \\
 This task is daunting because the probability and structural
feature of the entire spectrum of microstates sampled by proteins,
is not clearly known {[}31, 32{]}. The sensitivity of the ensemble
of microstates to changes in environmental conditions (i.e., pH, temperature,
pressure, ligand binding, and concentrations of osmolytes and denaturants)
is also not well understood either {[}32{]}. Most importantly, the
manner in which local fluctuations are coupled to larger, more global
structural transitions - isn't known either. Hence a mathematical
construct that attempts to model the situation must essentially be
top-down in its approach (to circumvent tackling the time dependent
couplings between each and every local fluctuations), yet extract
the necessary information regarding the emergence of any biophysical
property. Rather than assuming the aforementioned interatomic interactions
to be linear, we have resorted to model the situation from a nonlinear
perspective. Reasons behind this assertion and the relevance of it
in the present study are numerous. Protein fold in the crowded milieu
of the cell, where the density of protein is $\sim\left(0.2-0.3\right)$
g/mL, Hence the huge number of intramolecular interatomic interactions
(formation, breakage and reformation of hydrogen bonds, salt bridges
etc ..) take place in competition with similar intermolecular interatomic
interactions; which, of course, may be detrimental to the folding
process {[}33{]}. Despite it, the process usually yields the native
state in a matter of milliseconds to seconds. This inherent swiftness
of the process implies a series of ordered events or intermediates,
a fact that has intrigued researchers for several decades {[}34{]},{[}35{]}.
However, the order in which these intramolecular events take place
are not known in general terms, even today. Evolution of a biophysical
property from its mesoscopic limit to macroscopic limit (statistical
parametric limit of 32 atoms) can be studied from the implementation
of the present (general) mathematical framework. Careful (computational)
implementation of such mathematical framework can (possibly) resolve
the contradictions in protein folding/unfolding mechanisms (some models
to describe denatured state banks either on the sum of individual
amino acids {[}36,37{]} or on an extended, completely solvent-exposed
polypeptide chain {[}38,39{]}. This assumption is at odds with experimental
evidence showing that the denatured state in the absence of denaturants
is rather compact {[}40,41{]}), as elaborated in earlier {[}23{]}
study). Thus, the present study is not merely about theoretical pursuit
but has several practical uses too.\\

\section{\underbar{Methodology :}}

Application of control theoretic constructs to study biological systems
is not exactly common, but presence of such rigorous mathematical
constructs are found in many recent works. All of these works, in
some form or the other, demonstrate that reconstruction of some specific
regulated states under conditions of limited information - can be
achieved extremely efficiently through the control theoretic constructs.
While many of these (pioneering) deductions are applicable to paradigms
in systems biology {[}42-44{]}, isolated instances of insightful treatment
of genetic regulation can be found too {[}45{]}. Consideration of
the role of observer {[}46, 47{]} in an essentially nonlinear paradigm
of systemic description of biological systems is successfully achieved
in the later. However, control theoretic studies on protein biophysical
factors with similar rigorous standpoint, were not found. Several
works on control theoretic constructs attempt to model the systems
from a (time-invariant) linear perspective {[}48, 49{]}. Since such
formalism is not relevant in the attempts to describe biological systems,
the treatment of the same in the present work is attempted from time-dependent
perspective, completely. On the other hand, since we are attempting
to observe the emergence of protein biophysical features (in a mesoscopic
scale), the state-space oriented control studies (as have been attempted
in some biological paradigms {[}50, 51{]}) were not explicitly touched
in the present study. We note that possibility of application of control
theory in the context of protein structure prediction through NMR
was discussed in a previous article {[}52{]}, the actual conditions
of observability of the emergence of a (statistical) macroscopic property
was not obtained there.\\
 \\
 While the present study owes its philosophical basis to the aforementioned
studies (and many others {[}53, 54{]}) it differs from all of the
above; because, to our knowledge, this is the first attempt to propose
a theoretical framework that attempts to observe how the measurable
macroscopic biophysical properties of proteins come to being from
(microscopic) time-dependent and context-dependent interatomic interactions.\\
 \\
 Having established the reason behind such studies, here we embark
on derivation of the algorithm. This is done in two parts. In the
first section, the definition of protein from the perspective of control
theory is put into place. The next section then approaches the problem
in a step-by-step manner to deduce the conditions that will unambiguously
define whether any (sub)set of atoms of a protein is completely, or
partially or not observable with respect to a biophysical property.
Calculations for the (approximated) linear case, are also kept here;
because in certain particular cases the computational implementation
of the non-linear case may become difficult. The actual description
of the process, however, can only be found from derivations of the
non-linear case.\\
 \\
 \textbf{\underbar{Section - 1) : Definition of the system(a single
protein)}}\\
 \\
 \textbf{Case-1)} \textbf{General representation of protein interior
parameters with linear differential equation.}\\
 We approach to objectively describe these time-dependent and
context-dependent correlations amongst protein structural parameters
by representing any arbitrarily chosen protein with a linear differential
equation with sufficient capacity to describe the (time-dependent)
dynamic-dependencies of protein structural parameters on one another.
Such an approach provides us with a computationally implementable
simple framework with adequate rigor, given by :\\
 \\
 \begin{equation}
\dot{x}\left(t\right)=A(t)\, x(t)\;+\: f(t)\end{equation}
 \\
 where, $x$ is a $n-vector$, $A\left(t\right)$ is an $n\,\times\, n$
continuous matrix on an open interval $I$ in $R$, and $f\left(t\right)$
is locally square integrable on some arbitrarily chosen interval $\left(a,\, b\right)$,
viz. $f\left(t\right)\in L_{n}^{2}\,\left(\,\left[a,b\right]\,\right)$.
In other words, the space of all measurable $n-vector$ functions
$f(t)$ defined for $t\in\left[a,b\right]\,=\, J$ with values $f\left(t\right)\,\in\, R^{n}\,,\; t\in J$
such that $\int_{a}^{b}\,|\, f\left(t\right)|^{2}\, dt\,<\,\infty$.\\
 \\
 \\
 We can write $eq^{n}-1$, in the following equivalent form as
:\\
 \begin{equation}
x\left(t\right)\,=\, x_{0}\:+\:\int_{t_{0}}^{t}\,\left[A\left(s\right)x\left(s\right)\;+\; f\left(s\right)\right]ds\end{equation}
 \\
 Subsequently we can define the successive approximations by the
relations :\\
 $x_{0}\,\left(t\right)\:=\: x_{0}$\\
 and\\
 $x_{n+1}\left(t\right)\;=\; x_{0}\:+\;\int_{t_{0}}^{t}\,\left[A\left(s\right)x_{n}\left(s\right)\;+\; f\left(s\right)\right]ds,\quad t\in J,\quad n=0,1,2,\ldots\:$\\
 \\
 The solution of $eq^{n}-2$ with $x\left(t_{0}\right)\,=\, x_{0}$
is given by :\\
 \begin{equation}
x\left(t\right)\,=\, X\left(t,\, t_{0}\right)x_{0}\;+\;\int_{t_{0}}^{t}\, X\left(t,\, s\right)f\left(s\right)ds,\end{equation}
 \\
 where $X\left(t,\, t_{0}\right)$ is the fundamental matrix solution
of homogeneous equation, $\:\dot{x}\left(t\right)=A\left(t\right)x\left(t\right)\;$,\\
 which has the following properties :\\
 1) $X\left(t_{0},\, t_{0}\right)\:=\: I\:$(the identity matrix).
\\
 2) $X\left(t,\, t_{0}\right)\:=X\left(t,s\right)X\left(s,t_{0}\right),\quad\quad t_{0}\leq s\leq t$\\
 3) $X\left(t,\, s\right)\,=\, X^{-1}\,\left(s,\, t\right)$\\
 \\
 \\
 \textbf{Case-2) General representation of protein interior parameters
with non-linear differential equation.}\\
 Of course, a representation scheme similar to $eq^{n}-3$ can
be obtained for a non-linear differential equation of the form :\\
 \begin{equation}
\dot{x}\left(t\right)=A(t)\, x(t)\;+\: f(t,\, x)\end{equation}
 \\
 where $f\left(t,\, x\right)$ is continuous on $J\,\times\, R^{n}$.
For $eq^{n}-4$, the solution of $x\left(t\right)$ with $x\left(t_{0}\right)=x_{0}$
can be written as :\\
 \begin{equation}
x\left(t\right)\,=\, X\left(t,t_{0}\right)x_{0}\;+\;\int_{t_{0}}^{t}\, X\left(t,s\right)f\left(s,x\left(s\right)\right)ds,\end{equation}
 \\
 where $X\left(t,\, t_{0}\right)$ is the fundamental matrix solution
of homogeneous equation. With this non-linear representation, complexities
of solution increases undoubtedly, but under suitable conditions on
$A$ and $f$ , one can establish the existence of solution of the
non-linear equation (4). \\
 \\
 \\
 \textbf{\underbar{Section - 2) : Criteria of observability of
protein interior parameter dynamics.}}\\
 Based on the definition of the system (that is, an arbitrarily
chosen protein) as provided above we now proceed to derive the conditions
for observability of emergence of any biophysical property from any
arbitrarily chosen (sub)set of atoms belonging to the protein.\\
 \\
 \textbf{\underbar{Case-1) :}}\\
 \textbf{\underbar{Observability, under the assumption that proteins
are linear systems :}}\\
 We continue with our description of interactions of various structural
parameters in protein interior and the emerging property with the
differential equation, $\dot{x}\left(t\right)=A(t)\, x\;+\: f(t)$
(all the symbols retain their meaning from $eq^{n}-1$)\\
 \\
 Here, in section-2, we attempt to approach the description of
the process when $eq^{n}-1$ is subjected to a linear observation
process, described with simple form, viz:\\
 \begin{equation}
y\:=O\left(t\right)x\;+\;\hat{O}\left(t\right)f,\quad\; y\,\in\, R^{n}\end{equation}
 \\
 Assuming that the interaction of various structural parameters
was taking place in a time interval $\left[t_{0},\, t_{1}\right]\,\subset\,\left(a,b\right)$
and $x\left(t_{0}\right)=x_{0}\,\in R^{n}$, we had arrived at $eq^{n}-3$.
We start the derivation necessary to describe observability of the
protein with aforementioned structural parameters, by noticing first
admitting that observation of the relevant phenomenon under question
itself is a time-dependent process; and second, if $f$ is known function,
for example $f\left(t\right)=B\left(t\right)u\left(t\right)$ with
$u\left(t\right)$ being a control then, in principle the term $\hat{O}\left(t\right)f$
in $eq^{n}-6$ and $O\left(t\right)$ times the integral of $eq^{n}-3$
can be subtracted from :\\
 \\
 $y\left(t\right)=O\left(t\right)\, X\left(t,\, t_{0}\right)x_{0}\;+\: O\left(t\right)\,\int_{t_{0}}^{t}\, X\left(t,\, s\right)f\left(s\right)ds+\;\hat{O}\left(t\right)f\left(t\right)$\\
 \\
 to yield the modified closed form expression for observation
given by :\\
 \begin{equation}
\hat{y}\left(t\right)=O\left(t\right)\, X\left(t,\, t_{0}\right)x_{0}\end{equation}
 \\
 The term $X\left(t,t_{0}\right)x_{0}$ in $eq^{n}-7$ satisfies
the homogeneous equation\\
 \begin{equation}
\dot{x}=A\left(t\right)x\end{equation}
 \\
 Therefore the expression to represent linear observation on a
protein whose interior structural dependencies can be described by
a linear differential equation, is obtained as :\\
 \begin{equation}
y\left(t\right)=O\left(t\right)x\left(t\right)\end{equation}
 \\
 Hence, the original question about obtaining information about
a system described by $eq^{n}-1$ with the help of an observation
scheme described by $eq^{n}-6$, reduces to the same question for
the corresponding homogeneous system, described by $eq^{n}-8$ and
the homogeneous observation described by $eq^{n}-9$. This transformation
between paradigm of questions mark the change in modes of studies
because the present platform (comprised of $eq^{n}-8$ and $eq^{n}-9$)
offers us a homogeneity in the treatment of the problem in general;
something that wasn't ensured in the platform comprised of $eq^{n}-1$
and $eq^{n}-6$.\\
 \\
 However, to make meaningful predictive studies, the present framework
needs to be modified further regarding a suitable description scheme
to describe temporal frame of reference. Thus, without any loss of
generality, we perform the translation of the the origin, so that
$\tau=t\,-\, t_{0}$. This accounts for the limits $t_{0}\,\longrightarrow0$
, whereby $t_{1}\,\longrightarrow\left(t_{1}\,-\, t_{0}\right)=T.$\\
 \\
 \\
 \\
 \\
 To formalize the problem we define $observability$ in the manner
that, the system represented by $eq^{n}-8$ and $eq^{n}-9$ is observable
(that is, the pair $\left(O\left(t\right),\, A\left(t\right)\right)$
is observable) on the time interval $\left[0,T\right]$\\
 \textbf{iff} $\; y\left(t\right)=O\left(t\right)x\left(t\right)=0,\qquad t\in\left[0,T\right]\quad$
implies $x\left(t\right)=0,\quad t\in\left[0,T\right].$\\
 (which is equivalent to the assertion $x\left(0\right)=x_{0}=0$).
\\
 \\
 Thus, the re-defined version of the problem is to identify (and/or
develop) the conditions for observability on the matrices $A\left(t\right)$
and $O\left(t\right)$.\\
 \\
 We approach the problem by denoting the space of square integrable
$r-vector$ functions on $\left[0,T\right]$ by $L_{r}^{2}\,\left[0,T\right]$.\\
 \\
 At this point we propose theorem-1, theorem of 'connection between
independence of protein structural parameter and their corresponding
observability'.\\
 \\
 \textbf{\underbar{Theorem-1) :}} \\
 If the structural parameters corresponding to any protein can
be represented by vectors $\; x_{1},x_{2},\ldots,x_{k}\;$ in finite
dimensional Euclidean space $R^{n}$, and if $\; x_{1}\left(t\right),x_{2}\left(t\right),\ldots,x_{k}\left(t\right)\;$
be the corresponding solutions of $eq^{n}-8$ for them in $\:\left[0,T\right]\:$
with $\; x\left(0\right)=x_{i}\:,\: i=1,2,\ldots,k\,$; further if
the corresponding observations $y_{i}$ on $\:\left[0,T\right]\:$
can be defined by $\: y_{i}\left(t\right)=O\left(t\right)x_{i}\left(t\right)\:,\; t\in\left[0,T\right]$;
then the observed linear system described by $eq^{n}-8$ and $eq^{n}-9$
is observable on $\left[0,T\right]$; if and only if, $\: y_{i}\:$
are linearly independent in $L_{r}^{2}\,\left[0,T\right]$ whenever
the $\: x_{i}\:$ are linearly independent in the same finite dimensional
Euclidean space $R^{n}$.\\
 \\
 \textbf{\underbar{Proof :}}\\
 The solutions $x_{i}\left(t\right)$ are linearly independent
in $L_{r}^{2}\,\left[0,T\right]$ only in the case when $\: x_{i}\:$
are linearly independent in $R^{n}$. If $eq^{n}-8$ and $eq^{n}-9$
is observable and \\
 \begin{equation}
y\left(t\right)=\sum_{i=1}^{k}\, c_{i}y_{i}\left(t\right)=0\end{equation}
 \\
 then the corresponding solution also vanishes. In other words,
it implies :\\
 \begin{equation}
x\left(t\right)=\sum_{i=1}^{k}\, c_{i}x_{i}\left(t\right)=0\end{equation}
 \\
 and in particular \\
 \begin{equation}
\sum_{i=1}^{k}\, c_{i}x_{i}\left(0\right)=\sum_{i=1}^{k}\, c_{i}x_{i}=0\end{equation}
 \\
 In the case where $x_{i}$'s are linearly independent, we have
$c_{1}=c_{2}=\ldots=c_{k}=0$. Hence, from $eq^{n}-10$, we can conclude
that in such a case $y_{i}$'s will be linearly independent too.\\
 \\
 On the other hand (evidently, in the more $general\: case$)
if there exists linearly independent $\; x_{1},x_{2},\ldots,x_{k}\;$
such that the associated observations of them, namely $\; y_{1}\left(t\right),y_{2}\left(t\right),\ldots,y_{k}\left(t\right)\;$
are $not$ independent (that is, are dependent on $L_{r}^{2}\,\left[0,T\right]$),
then letting $\; c_{1}=c_{2}=\ldots=c_{k}\:$ are $\, not\,$ all
zero, such that \\
 $y\left(t\right)=\sum_{i=1}^{k}\, c_{i}y_{i}\left(t\right)=0$,\\
 we notice that $\; y\left(t\right)\;$ becomes an identically
vanishing observation on the solution $\; x\left(t\right)=\sum_{i=1}^{k}\, c_{i}x_{i}\left(t\right)\;$,
which is $\, not\,$ the zero solution of $eq^{n}-8$, because in
such a case, $\; x_{1}=x_{1}\left(0\right),\, x_{2}=x_{2}\left(0\right),\,\ldots\,,\, x_{k}=x_{k}\left(0\right)\:$
$\, are\,$ linearly independent.\\
 \\
 Hence, in such a case, we can conclude that $eq^{n}-8$ and $eq^{n}-9$
will not be observable. \textbf{Q.E.D}\\
 \\
 Proof of theorem-1 ('connection between independence of protein
structural parameter and their corresponding observability') paves
the way for a more general theorem, the theorem-2, 'observability
of protein structural parameters as components of a linear system'\\
 \\
 \\
 \\
 \textbf{\underbar{Theorem-2) :}}\\
 The system(protein), described by $eq^{n}-8$ and $eq^{n}-9$,
is observable on time interval $\left[0,T\right]$ iff the observability
Grammian matrix, given by :\\
 $\Phi\left(0,T\right)=\int_{0}^{T}\, X^{*}\left(t,0\right)O^{*}\left(t\right)O\left(t\right)X\left(t,0\right)dt$\\
 (where $X^{*}\left(t,0\right)$ and $O^{*}\left(t\right)$ are
the transposes of $X\left(t,0\right)$ and $O\left(t\right)$) is
positive definite.\\
 \\
 \textbf{\underbar{Proof :}}\\
 The solution of $x\left(t\right)$ of $eq^{n}-8$ corresponding
to the initial condition $x\left(0\right)=x_{0}$ is given by :\\
 $x\left(t\right)=X\left(t,0\right)x_{0}$\\
 and we obtain $y\left(t\right)=O\left(t\right)x\left(t\right)=O\left(t\right)X\left(t,0\right)x_{0}$\\
 \\
 \begin{eqnarray*}
\parallel y\parallel^{2} & = & \int_{0}^{T}\, y^{*}\left(t\right)y\left(t\right)dt\\
 & = & x_{0}^{*}\,\int_{0}^{T}\, X\left(t,0\right)O^{*}\left(t\right)O\left(t\right)X\left(t,0\right)dt\, x_{0}\\
 & = & x_{0}^{*}\,\Phi\left(0,T\right)\, x_{0}\end{eqnarray*}
 \\
 - a quadratic form in $x_{0}$.\\
 Clearly $\Phi\left(0,T\right)$ is a symmetric $\, n\times n\:$
matrix.\\
 \\
 If $\Phi\left(0,T\right)$ is positive definite then :\\
 $y=0\quad\Rightarrow x_{0}^{*}\,\Phi\left(0,T\right)\, x_{0}=0\quad\Rightarrow x_{0}=0$\\
 \\
 and then the system described by $eq^{n}-8$ and $eq^{n}-9$
is observable on $\left[0,T\right]$. If, $\Phi\left(0,T\right)$
is $\, not\,$ positive definite then it implies that there exists
some $x_{0}\neq0$ such that $x_{0}^{*}\,\Phi\left(0,T\right)\, x_{0}=0$.\\
 In such a case, $x\left(t\right)=X\left(t,0\right)\: x_{0}\neq0$
for $t\in\left[0,T\right]$ but since $\,\parallel y\parallel^{2}=0\,$,
it implies $\, y=0.\,$\\
 And therefore, we can conclude that system described by $eq^{n}-8$
and $eq^{n}-9$ is not observable on $\left[0,T\right]$.$\qquad\qquad$
\textbf{Q.E.D}\\
 \\
 \\
 \textbf{\underbar{Corollary :}} \\
 If the system described by $eq^{n}-8$ and $eq^{n}-9$ is observable
on $\left[s,t\right]$ then it is also observable on any interval
$\left[0,T\right]$ such that $0\leq s\leq t\leq T$.\\
 Proof of corollary :\\
 We have :\\
 \begin{eqnarray*}
\Phi\left(0,T\right) & = & X^{*}\left(s,0\right)\int_{0}^{T}\, X^{*}\left(\tau,s\right)O^{*}\left(\tau\right)O\left(\tau\right)X\left(\tau,s\right)d\tau\, X\left(s,0\right)\\
 & \geq & X^{*}\left(s,0\right)\Phi\left(s,t\right)X\left(s,0\right)\\
 & > & 0\end{eqnarray*}
 \\
 Hence the proof.\\
 \\
 \\
 \\
 \textbf{\underbar{Case-2) :}}\\
 \textbf{\underbar{Observability, under the assumption that proteins
are non-linear systems :}}\\
 In certain cases the dependencies amongst structural parameters
within any protein might not be governed by equations with simple
linear dependencies. This is probable too, considering that interactions
amongst protein structural determinants are time-dependent and context-dependent.
Hence, in such a case, without resorting to the linear (simplistic
and approximated) case, we will have to describe proteins as non-linear
systems. Here, instead of referring to $eq^{n}-1$, we start our descriptions
with $eq^{n}-4$, viz : $\dot{\: x}\left(t\right)=A(t)\, x(t)\;+\: f(t,\, x)$,
\\
 where $x$ and $f$ are $n-vectors$, $\: t\in I\:$, the real
time interval with linear observation $\: y=O\left(t\right)x\left(t\right)\:$,
where $y$ is an $m-vector$ $\left(m<n\right)$ and $A\left(t\right)$,
$f\left(t,\, x\right),$ $O\left(t\right)$ - are continuous with
respect to their arguments.\\
 \\
 Here, although admitting that case of non-linear representation
scheme is more general (and perhaps, more accurate), we resort mathematically
to describe the case as a linear system $\left(eq^{n}-4\right)$ with
perturbation $\, f\left(t,\, x\right)$. We assume further that the
it is feasible to describe the situation as one where $\left(eq^{n}-4\right)$
is being observed by a quantity $y$. In such a framework of description,
the problem of observability of $\left(eq^{n}-4\right)$ can be formulated
as one, where :\\
 it is required to find an unknown state at the present time $t$
from the quantity $y$, over the interval $\left[\theta,\, t\right]$
where $\theta$ denotes some past time, because, since $\,\left(m<n\right)\,$,
the equation $\: y=O\left(t\right)x\left(t\right)\:$ does not allow
immediate finding of $x$ and $y$.\\
 \\
 At this point, having loosely describing the framework to describe
the situation, we proceed to formally define the system as :\\
 \textbf{\underbar{NL-Defn-1)}} The system can be defined to be
$\, observable\,$ at time $t$, if there exists $\theta<t$ such
that the state of the system at time $t$, can be identified from
the knowledge of the system output over the interval $\,\left[\theta,t\right]$.
\\
 \textbf{\underbar{NL-Defn-2)}} If the system is observable at
every $\: t\in I\:$, it can be called 'completely observable'.\\
 \textbf{\underbar{NL-Defn-3)}} If the interval of output observation
mentioned in NL-Defn-1, can be made arbitrarily small, we speak of
differential observability over those intervals. \\
 \\
 We start our analysis of non-linear description of protein interior
by assuming that $\left(eq^{n}-4\right)$ has a unique solution for
any initial condition. If we denote $\tau$ as $\,\theta<\tau<t,\:$
the solution for $\left(eq^{n}-4\right)$ can be asserted to be uniquely
defined for $x=x\left(\tau\right)$ as the initial condition and (drawing
from aforementioned non-linear description of proteins in case-2 of
section-1) given by :\\
 $x\left(t\right)\,=\, X\left(t,\tau\right)x\left(\tau\right)\;+\;\int_{\tau}^{t}\, X\left(t,s\right)f\left(s,x\left(s\right)\right)ds$\\
 \\
 However, since the fundamental matrix is invertible in nature,
we have :\\
 \begin{equation}
x\left(t\right)\,=\, X\left(\tau,t\right)x\left(t\right)\;-\;\int_{\tau}^{t}\, X\left(\tau,s\right)f\left(s,x\left(s\right)\right)ds\end{equation}
 \\
 Correspondingly the $y\left(\tau\right)$ will be given by :\\
 \begin{equation}
y\left(\tau\right)=O\left(\tau\right)\, X\left(\tau,t\right)x\left(t\right)\;-\; O\left(\tau\right)\int_{\tau}^{t}\, X\left(\tau,s\right)f\left(s,x\left(s\right)\right)ds\end{equation}
 \\
 Describing the transpose of any matrix, with a star symbol, with
a little bit of rearrangement (by multiplying $eq^{n}-14$ with $X^{*}\left(\tau,t\right)O^{*}\left(\tau\right)$
from the left and integrating within the interval $\theta$ to $t$),
we obtain :\\
 \begin{eqnarray*}
\int_{\theta}^{t}X^{*}\left(\tau,t\right)O^{*}\left(\tau\right)y\left(\tau\right)d\tau & = & \left[\int_{\theta}^{t}X\left(\tau,t\right)O^{*}\left(\tau\right)O\left(\tau\right)X\left(\tau,t\right)drx\left(t\right)\right]\\
 & - & \left[\int_{\theta}^{t}X\left(\tau,t\right)O^{*}\left(\tau\right)O\left(\tau\right)\int_{\tau}^{t}\, X\left(\tau,s\right)f\left(s,x\left(s\right)\right)dsdr\right]\\
 & = & \Phi\left(\theta,t\right)x\left(t\right)\\
 & - & \left[\int_{\theta}^{t}\, X^{*}\left(s,t\right)\int_{\theta}^{s}\, X^{*}\left(\tau,s\right)O^{*}\left(\tau\right)O\left(\tau\right)X\left(\tau,s\right)d\tau f\left(s,x\left(s\right)\right)ds\right]\\
 & = & \Phi\left(\theta,t\right)x\left(t\right)-\left[\int_{\theta}^{t}\, X^{*}\left(s,t\right)\Phi\left(\theta,s\right)f\left(s,x\left(s\right)\right)ds\right]\end{eqnarray*}
 \\
 If the matrix $\Phi\left(\theta,t\right)$ is invertible, that
is, for a truncated linear system, \\
 \begin{equation}
\dot{x}=A\left(t\right)x,\qquad y\left(t\right)=O\left(t\right)x\end{equation}
 \\
 is observable; then from the last equation of the array of equations,\\
 $(\;\int_{\theta}^{t}X^{*}\left(\tau,t\right)O^{*}\left(\tau\right)y\left(\tau\right)d\tau=\Phi\left(\theta,t\right)x\left(t\right)-\left[\int_{\theta}^{t}\, X^{*}\left(s,t\right)\Phi\left(\theta,s\right)f\left(s,x\left(s\right)\right)ds\right]\;)$\\
 we obtain :\\
 $x\left(t\right)=\Phi^{-1}\left(\theta,t\right)\,\int_{\theta}^{t}\, X^{*}\left(s,t\right)O^{*}\left(s\right)y\left(s\right)ds\:+\:\Phi^{-1}\left(\theta,t\right)\,\int_{\theta}^{t}\, X^{*}\left(s,t\right)\Phi\left(\theta,s\right)f\left(s,x\left(s\right)\right)ds$\\
 \\
 If we assign :\\
 $U_{1}\left(t,\theta,s\right)=\Phi^{-1}\left(\theta,t\right)\,\int_{\theta}^{t}\, X^{*}\left(s,t\right)O^{*}\left(s\right)$\\
 and\\
 $U_{2}\left(t,\theta,s\right)=\Phi^{-1}\left(\theta,t\right)\,\int_{\theta}^{t}\, X^{*}\left(s,t\right)\Phi\left(\theta,s\right)$\\
 \\
 then the following compact relation can be obtained :\\
 \begin{equation}
x\left(t\right)=\int_{\theta}^{t}\, U_{1}\left(t,\theta,s\right)y\left(s\right)ds\;+\;\int_{\theta}^{t}\, U_{2}\left(t,\theta,s\right)f\left(s,x\left(s\right)\right)ds\end{equation}
 \\
 Equation-16 represents the relation of the unknown state x with
the observed output y over he interval $\left[\theta,t\right]$. \\
 \\
 In $eq^{n}-16$ the time $\theta$ may not be necessarily fixed,
and therefore $\theta$ can be replaced by $\tau$. Upon carrying
out this change, $eq^{n}-16$ can be substituted into $eq^{n}-13$,
and we obtain :\\
 \\
 \begin{equation}
x\left(\tau\right)=X\left(\tau,t\right)\,\int_{\tau}^{t}\, U_{1}\left(t,\tau,s\right)y\left(s\right)ds+X\left(\tau,t\right)\,\int_{\tau}^{t}\, U_{2}\left(t,\tau,s\right)f\left(s,x\left(s\right)\right)ds-\int_{\tau}^{t}\, X\left(\tau,s\right)f\left(s.x\left(s\right)\right)ds\end{equation}
 \\
 \\
 \\
 In compact form,\\
 \begin{equation}
x\left(\tau\right)=\int_{\tau}^{t}\, U_{3}\left(t,\tau,s\right)y\left(s\right)ds+\int_{\tau}^{t}\, U_{4}\left(t,\tau,s\right)f\left(s,x\left(s\right)\right)ds\quad for\left(\tau<t\right)\end{equation}
 \\
 where $\: U_{3}\left(t,\tau,s\right)=X\left(\tau,t\right)\, U_{1}\left(t,\tau,s\right)\;$,
\\
 and $\: U_{4}\left(t,\tau,s\right)=X\left(\tau,t\right)\, U_{2}\left(t,\tau,s\right)\,-\, X\left(\tau,s\right)\;$
\\
 \\
 On the basis of the derivations above we can put forward a proposition
as\\
 \textbf{\underbar{NL-Proposition-1) :}} \\
 Under the condition that the system (protein interior) is describable
by the differential equations $\dot{x}\left(t\right)=A(t)\, x(t)\;+\: f(t,\, x)$
and $y\left(t\right)=O\left(t\right)x$, it is globally \\
 $\mathbf{\, a)\,}$ observable at any time instance $t$, \\
 \textbf{$\,\mathbf{b)}\,$} completely observable, or \\
 \textbf{$\,\mathbf{c)}\,$} differentially observable, if the
following conditions hold \textbf{:}\\
 \textbf{1)} there exists a constant $c>0$, such that $\: det\,\Phi\left(t,\theta\right)\,\geq\, c\;$.\\
 \textbf{2)} $Eq^{n}-16$ has a unique solution for any $y$,
which is continuous on $\left[\theta,t\right]$ $\mathbf{\, a)\,}$
for some $\theta<t$, in case of an observable system at time $t$,
$\mathbf{\, b)\,}$ for all $t$ and for some $\theta<t$, in the
case of a completely observable system, or $\mathbf{\, c)\,}$ for
all $t$ and for all $\theta<t$, in the case of a differentially
observable system\\
 \\
 \\
 A careful of reading of the description of the situation suggests
that $eq^{n}-16$ in 'NL-proposition-1' can be replaced by $eq^{n}-17$.
In the case of such replacement the same results will be valid, but
with some simple change of variables. However the question that whether
$eq^{n}-16$ or $eq^{n}-17$ has a unique solution is difficult to
evaluate. The difficulty arises because we notice that if the interval
of integration in $eq^{n}-16$ or $eq^{n}-17$ is suitably changed,
$eq^{n}-16$ or $eq^{n}-17$ may then be considered as a nonlinear
operator equation on a continuous function space. Thus, we will have
to resort to Banach's contraction mapping theorem to these nonlinear
equations. \\
 \\
 Without venturing into the general case, we consider a special
system described by :\\
 \begin{equation}
\dot{x}=A\left(t\right)x\;+\;\epsilon f\left(t,x\right)\end{equation}
 \begin{equation}
y=O\left(t\right)x\end{equation}
 \\
 where, $\epsilon$ is a scaler positive constant. We assume that
the following condition is satisfied :\\
 \begin{equation}
\parallel f\left(t,x_{1}\right)-f\left(t,x_{2}\right)\parallel\,\leq\, K\parallel x_{1}-x_{2}\parallel,\quad\left(K\geq0\right)\end{equation}
 \\
 A general solution $x\left(t\right)$ for $eq^{n}-18$ with $x=x\left(\tau\right)$
as a formal initial condition is:\\
 \begin{equation}
x\left(\tau\right)=X\left(\tau,t\right)x\left(t\right)\,-\,\epsilon\int_{\tau}^{t}\, X\left(\tau,t\right)f\left(s,x\left(s\right)\right)ds\end{equation}
 \\
 \\
 We create an analogue of $eq^{n}-16$ derivation from $eq^{n}-13$,
by starting with $eq^{n}-22$.\\
 Therefore :\\
 \begin{equation}
x\left(t\right)=\Phi^{-1}\left(\theta,t\right)\int_{\theta}^{t}X^{*}\left(s,t\right)O^{*}\left(s\right)y\left(s\right)ds+\epsilon\Phi^{-1}\left(\theta,t\right)\int_{\theta}^{t}X^{*}\left(s,t\right)\Phi\left(s,\theta\right)f\left(s,x\left(s\right)\right)ds\end{equation}
 \\
 \\
 Substituting $eq^{n}-22$ on $eq^{n}-21$, we obtain :\\
 \begin{eqnarray*}
x\left(\tau\right) & = & X\left(\tau,t\right)\Phi^{-1}\left(t,\theta\right)\,\int_{\theta}^{t}\, X^{*}\left(s,t\right)O^{*}\left(s\right)y\left(s\right)ds\\
 & + & \epsilon X\left(\tau,t\right)\Phi^{-1}\left(\theta,t\right)\,\int_{\theta}^{t}\, X^{*}\left(s,t\right)\Phi\left(s,\theta\right)f\left(s,x\left(s\right)\right)ds\\
 & - & \epsilon\int_{\tau}^{t}\, X^{*}\left(\tau,s\right)f\left(s,x\left(s\right)\right)ds\quad\left(\theta\leq\tau\leq t\right)\end{eqnarray*}
 \\
 We denote this condition as 'NL-cond-1'.\\
 \\
 Consequently, in order for the system described by $eq^{n}-18$
and $eq^{n}-19$, to be observable, it is sufficient that the inverse
of $\Phi\left(t,\theta\right)$ exists, the solution of $eq^{n}-23$
exists and it is unique.\\
 \\
 At this point, if we assume that there exists solution of $x_{1},x_{2}\:\left(x_{1}\neq x_{2}\right)$
of $eq^{n}-23$ for a given $y$, then, using $eq^{n}-20$, we obtain
:\\
 \begin{eqnarray*}
\left(|x_{1}\,\left(\tau\right)-x_{2}\,\left(\tau\right)\right) & \leq & \epsilon\int_{\tau}^{t}|\, X\left(\tau,s\right)|K|x_{1}\left(s\right)-x_{2}\left(s\right)|ds\\
 & + & \epsilon|X\left(\tau,t\right)\Phi^{-1}\left(t,\theta\right)|\int_{\theta}^{t}|X^{*}\left(s,t\right)\Phi\left(s,\theta\right)|K|x_{1}\left(s\right)-x_{2}\left(s\right)|ds\\
 & \leq & \epsilon k_{1}\left(t,\theta\right)\left(t-\tau\right)\parallel x_{1}-x_{2}\parallel+\epsilon k_{2}\left(t,\theta\right)\parallel x_{1}-x_{2}\parallel\left(t-\theta\right)\end{eqnarray*}
 \\
 where\\
 $k_{1}\left(t-\theta\right)=\underset{\theta<\tau<s<t}{max}|X\left(\tau,t\right)\Phi^{-1}\left(t,\theta\right)||X^{*}\left(s,t\right)\Phi\left(s,\theta\right)K$\\
 \\
 From this, there exists a $k\left(t,\theta\right)$ such that
:\\
 \begin{equation}
\parallel x_{1}-x_{2}\parallel\leq\epsilon k\left(t,\theta\right)\left(t-\theta\right)\parallel x_{1}-x_{2}\parallel\end{equation}
 \\
 where $k\left(t,\theta\right)=k_{1}\left(t,\theta\right)+k_{2}\left(t,\theta\right)$
\\
 \\
 Hence, most importantly, if $\epsilon$ satisfies the inequality
:\\
 \begin{equation}
\epsilon k\left(t,\theta\right)\left(t-\theta\right)<1\end{equation}
 \\
 it follows that $x_{1}=x_{2}$ on $\left[\theta,t\right]$.\\
 \\
 This contradiction leads to the next proposition for a sufficient
condition for the observability of the system described by $eq^{n}-18$
and $eq^{n}-19$, since the condition, 'NL-cond-1' necessarily guarantees
the existence of solutions of $eq^{n}-23$.\\
 \\
 \\
 Hence, finally we have :\\
 \textbf{\underbar{NL-Proposition-2) :}} \\
 The system described by $eq^{n}-18$ and $eq^{n}-19$, is globally
$\mathbf{\, a)\,}$ observable at the instance $t$, \textbf{$\,\mathbf{b)}\,$}
completely observable or \textbf{$\,\mathbf{c)}\,$} differentially
observable, if the following conditions hold \textbf{:}\\
 \textbf{1)} there exists a constant $c>0$, such that $\: det\,\Phi\left(t,\theta\right)\,\geq\, c\;$.\\
 \textbf{2)} a positive constant, $\epsilon$ , satisfies \\
 $\epsilon<1/k\left(t,\theta\right)\left(t-\theta\right)$\\
 $\mathbf{\, a)\,}$ for some $\theta<t$, in case of an observable
system at time $t$, \\
 $\mathbf{\, b)\,}$ for all $t$ and for some $\theta<t$, in
the case of a completely observable system, and\\
 $\mathbf{\, c)\,}$ for all $t$ and for all $\theta<t$, in
the case of a differentially observable system.\\
 \\

\section{\underbar{Results and Discussion :}}

\textbf{\underbar{(Applicability of the algorithm on four different
spheres in}} \\
 \textbf{\underbar{Protein Biophysics) :}} \\
The present algorithm proposes a rigorous and reliable template
for a series of algorithms that can be constructed to study the emergence
of various biophysical factors. However, the actual implementation
of these ideas might require superlative computational facilities
that are not easily available in contemporary scenario; though the
possibility of using such facilities in near future seems genuine.
Due to prohibitive computational cost, actual implementation of these
algorithms could not be achieved in the present work. In the absence
of obtained data, in this section we present exact schemes to study
the emergence of actual biophysical properties from the mathematical
discourse that has been presented. Out of innumerable possibilities
of application of this algorithm, we chose four paradigms; on which,
the present study can (tangibly) be enormously impacting. In each
of these (extremely well-studied) spheres of protein biophysics, the
applicability of the present algorithm is clearly mentioned, alongside
the utilitarian benefits that application of this algorithm can provide
it with.\\
 \\
 \\
 \\
 \textbf{\underbar{Applicability - 1) Case of Hydrophobicity :}}\\
 \textbf{\underbar{1.1) Scope of applicability of the present
algorithm :}}\\
 Origin of hydrophobicity from a bottom-up approach has long been
a subject of fascination and enormous debate amongst physicists and
chemists for the last thirty years. A summary of this entire spectrum
of views, debates and scopes of (possible) confusions regarding various
ways of defining what hydrophobicity is, what solvation is, the origin
of hydrophobicity, (possible) correlation between hydrophobicity and
polarizability in different molecules under different boundary conditions;
are all well-documented {[}55{]}. Astonishment at the lack of our
understanding of the molecular mechanism causing hydrophobicity, as
has been expressed in a recent work {[}56{]}, is therefore justified.
However, it is neither in the scope nor in the motivation of the present
work to reflect upon these opinions. We start our argument by noting
down the pattern that the origin of hydrophobicity can be traced back
to some kind of inter-atomic interactions, was never questioned by
any of the proposed theories. These inter-atomic interactions are
bound to cause certain fluctuations in the spatial coordinate of the
atoms, is obvious too. The present work, attempts to derive the conditions
by which information about the extent of fluctuations in spatial coordinate
of a subset of atoms can be inferred from macroscopically measured
parameters. For example, if one hypothesizes that origin of hydrophobicity
can be studied from collective effects of dispersion forces in the
interior and exterior of the proteins, he (she, at any rate) can find
out the precise number of inter-atomic interactions that account for
the emergence of hydrophobicity. More interestingly, the starting
point of his study will not involve any guesswork concerning the number
of atoms; but will be some macroscopic measure of hydrophobicity (say,
hydrophobic fractal dimension {[}19{]}, or hydrophobic moment of a
protein {[}57{]}) and the number of atoms giving rise to it, will
be obtained as the result. Categorically, the user needs to assign
the measured (computationally or experimentally) change of (cumulative)
hydrophobicity content (as provided by the constructs proposed in
{[}19{]} or {[}57{]}) due to any set of atoms, to the variable $\dot{\, x\,}$;
the contact-map information for these atoms in the matrix $\, A\,$;
and the dependency of hydrophobicity on (quantum mechanically calculated)
partial charges or any other parameter that the user considers necessary
to describe dispersion related effects, in the (context-dependent)
function $\, f\left(x,\, t\right)\:$, of the $eq^{n}-4$. Or alternatively,
he can input the (time-dependent) coordinate information of the relevant
set of atoms in the variable $\, x\,$ and keeping the rest of the
parameters invariant, attempt to measure the change of hydrophobicity
content by scrutinizing the magnitude of $\dot{\, x\,}$. Owing to
the high degree of flexibility inherent in the algorithm, the effectiveness
of the proposed scheme lies primarily with the discretion of the user;
because that ensures the appropriate choice of parameters to be assigned
to the function $\, f\left(x,\, t\right)\:$ (as an example, instead
of resorting to (subjective) choice of parameters that might provide
an appropriate framework to study effect of dispersion related forces
on hydrophobicity, if the user had used atomic hydrophobicity magnitudes
{[}120{]} for every atom, the change in the content of hydrophobicity
could have been measured differently). Utility of the present scheme
therefore lies in the fact that it can study unambiguously how exactly
from the nano-scale (small number of distinguishable atoms), the macroscopic
property of thermodynamic nature (hydrophobicity) is emerging. Putting
the proposition differently, the algorithm can predict precisely how
much of the inter-atomic fluctuations will be observable with the
known (macroscopic) index of hydrophobicity. Posing the same question
in still different words, the present methodology can be used to ascertain
the precise lower limit of the number of atoms which are necessary
to observe emergence of the property named hydrophobicity. The mathematical
construct presented here is general. It can be put to rigorous use
by suitably considering only the pertinent aspect of it (according
to the nature of requirement of any particular problem) where it might
find potential utility. \\
 \\
 Such categorical information about the number and character of
atom-cluster that produces hydrophobicity due to their inter-atomic
interactions, becomes indispensable in order to probe recent questions
regarding the nature of hydrophobicity. For example, a steady flow
of opinions could be recorded over the last decade, which argued that
hydrophobic effect is not necessarily an entropic phenomenon; it can
be enthalpic or entropic depending on the temperature and the geometric
characteristics of the solute{[}58-61{]}. The difficulty with attempting
this problem stems primarily from the inherent contradiction, namely,
geometrical descriptions (distinguishable object based) and thermodynamic
(statistical) descriptions work at two different levels of systemic
descriptions. While the former is primarily bottom-up (nano scaled)
in its nature the later is top-down (macroscopic). The present theory
provides a quantitative tool-set to examine the emerging properties
in their nascent form in mesoscopic scale.\\
 \\
 \textbf{\underbar{1.2) Scope of applicability of the present
algorithm :}}\\
 It has been found experimentally that the relationship between
{}``bulk hydrophobic interaction'', exposure of hydrophobic residues
from its pure phase to water and {}``pair hydrophobic interaction''
potential of mean force (PMF) in water is nonlinear{[}62, 63{]}. Although
various experimental studies from varying perspectives (studies related
to virial coefficients, Kirkwood-Buff integrals, and on related spatially
integrated quantities {[}62-65{]}), have studied the nature of hydrophobicity
and many have attempted to focus purely on the multiple facets of
PMA {[}66-68{]}, the spatial dependence of PMF itself through to direct
experimental mechanism is difficult to obtain. Here, to know the precise
nature of spatial extent of PMF, rather than resorting to the simulation-centric
studies, we can resort to the rigorous mathematical treatise presented
here. By describing the fluctuating magnitude of temperature dependent
spatial extent of PMF with the variable $\dot{\, x\,}$, the (time-dependent)
contact map with matrix $\, A\,$, and the expression for dependency
of exposure of hydrophobic residues from its pure phase to water on
the (time-dependent and context-dependent) {}``pair hydrophobic interaction''
PMF of residues in $eq^{n}-4$, a consistent scheme can be constructed
to solve for the spatial extent of PMF for the concerned set of residues.
As has been mentioned before, the scheme constructed in this work
is a general one and the effectiveness of its depends crucially on
the judicious choice of parameters that are assigned to various terms
in $eq^{n}-4$ (or in $eq^{n}-1$, if the description is linear).\\
 \\
 \\
 \textbf{\underbar{Applicability - 2) Case of polarizability :}}\\
 \textbf{\underbar{2.1) Scope of applicability of the present
algorithm :}}\\
 A low polarizability in the interior of the protein implies a
low magnitude of dielectric constant, which in turn implies a conducive
environment for electrostatic interactions. Detailed account of electrostatic
interactions of ionized side chains is necessary requirement for serious
examination of protein stability because they are comparatively long-ranged
ones amongst biophysical forces{[}69{]}. All pH-dependent properties
of proteins are (predictably) governed by the electrostatic interactions
between ionizable side chains. Owing to this coupling to chemical
protonation equilibria, protein electrostatics can be probed directly
through measurements of pKa values {[}70-74{]}. The effect of electrostatic
interactions is usually quantified in terms of the shift $\,\Delta pK_{a}$,
of the pKa value of an ionizable group in a protein relative to the
pKa values of the same group in a small reference molecule in dilute
aqueous solution. \\
 \\
 Many aspects of protein pKa shifts are known to us. However,
solution to the basic inverse problem \textbf{{[}P-1{]}}, viz., given
a particular magnitude of $\,\Delta pK_{a}$, for residues either
in the surface or in the interior, what should be the minimum number
of residues, which might produces it ? - is not easily obtainable
in a general sense. Furthermore, despite immense efforts, computational
and/or theoretical approaches that can reliably predict the large
pKa shifts observed for buried residues in a general sense - remains
difficult to find. While one aspect of these problems lie in the computational
problems while considering ionization-induced water penetration and
conformational changes in pKa calculations, the other aspect points
to the lack of a theoretically sound scheme that can describe the
emergence of a macroscopic property from its inception to the subsequent
phases, as the number of residues that contribute to produce the property
increases over time. Substituting the pKa values for every amino acid
(in a sub-set of amino acids under consideration) in variable $x$,
incorporating the contact-map information in the matrix $A$, and
describing desolvation or conformational fluctuations (or any other
parameter that the user thinks necessary) in the context-dependent
(non-linear) function $f$ in the $eq^{n}-4$, one can attempt to
obtain a quantitative magnitude for $\,\Delta pK_{a}$ as the output
$\dot{\, x\,}$. The need for such a thorough scheme becomes even
acute when one attempts to delve into the uniform dielectric continuum
model of protein ineterior electrostatics. In such a model, the entire
effect due to polarizability is described through a single dielectric
constant (DC). (As a result, electrostatically highly heterogeneous{[}69{]}
and anisotropic {[}75{]} protein interior is represented through an
overtly simplified construct. The shortcomings of such model have
been commented upon by many {[}69,72,76{]}. The magnitude of DC becomes
a complex function of the extents to which formal charges, partial
charges, and dipoles {[}72{]} are considered. The effective DC is
calculated from the response of the entire protein to an externally
applied electric field; which in its turn, is calculated from the
total dipole moment fluctuation of the protein, through computational
methods. However, it has been reported by many that the magnitude
of dipole moment fluctuation is significantly affected by charged
surface residues {[}77-79{]}. Hence, to what accuracy will such a
construct be taking into account the self-energy of deeply buried
ionizable residues, - remains unclear. Having said that, many a studies
over the years have (successfully) addressed various aspects of the
aforementioned complex issues. Rigorous computational examinations
of the pKa shifts have been undertaken from the framework of macroscopic
dielectric continuum models, via semi-macroscopic partial charge {[}79,
80{]}, lattice dipole {[}72{]} models, and finally to all-atom simulations
{[}74{]}. However, answer to another basic and general inverse problem
\textbf{(P-2)}, namely, electrostatic effects due to how many buried
residues are being reflected in a measured magnitude of dipole moment
fluctuation of the protein - could not be found from the purview of
the aforementioned studies. \\
 \\
 The commonality in \textbf{P-1} and \textbf{P-2} is striking.
Both of them are asking extremely basic questions. Computational constructs,
regardless of how sophisticated they are, cannot provide the answers
to them. Instead, a thorough mathematical model, which treats protein
interior as a fluctuating, nonlinear (DC behaves in a nonlinear manner
{[}81{]}) and time-dependent system, might help us in finding the
answers to these basic (inverse problems). The mathematical model
presented in the present work attempts to achieve precisely the same.\\
 \\
 \textbf{\underbar{2.2) Scope of applicability of the present
algorithm :}}\\
 Importance of identifying the lower-threshold number of atoms
to understand the origin and nature of biophysical forces, is multifaceted.
We elaborate it in the context of studying the protein-protein interaction.
This complex process is driven primarily by the hydrophobic effect
and van der Waals interactions, with significant contribution from
entropy and electrostatics. However, a general criteria to understand
the precise extent of contribution of these, under varying biological
contexts, is extremely difficult to evaluate. Examination of the roles
of various components of electrostatic interactions during protein-protein
interaction assumes an inherent difficulty because of their (non-linear)
dependence on pH level and salt concentration {[}82{]}. To this end,
the finding that salt dependence of the binding is not correlated
with macroscopic parameters of the monomers {[}83,84{]}, merely serves
to underline the importance of categorical identification of the number
of atoms necessary to account for the origin of 'macroscopic' properties.
Furthermore, description of the process acquires a new level of difficulty
when one notices the latest finding {[}76{]} that, homo-complexes
and hetero-complexes adopt perfectly opposite scheme of electrostatics
during their formation. (For homo-complexes, contrary to intuitive
notions, in majority of the cases, the electrostatics opposes binding;
whereas, for hetero-complexes, it is somewhat like the role salt bridges
on protein stability; i.e., in some cases, it will favor the binding,
while in some other, it will oppose the binding.) Although this apparent
contradiction can be resolved to some extent by analyzing the charged-residue
density in the interface for the two classes{[}76{]}, construction
of a general scheme to describe the electrostatics of protein-protein
interaction. In the absence of a theoretically derived condition that
unambiguously demarcate the origin of several components of electrostatic
contributions in a generalized way, it has been reported that in certain
cases electrostatic energy favors binding, while in some other cases,
it opposes binding {[}76{]}. Furthermore, despite the consensus on
significance of specific pair-wise electrostatic interactions across
the interface {[}85-91{]}, the conclusions about the role of electrostatics
on binding affinity remain controversial. Having said that, we assert
that this confusion over the precise extent of comparative contribution
to various aspects of electrostatic forces on binding affinity can
be resolved objectively. Since all of the aforementioned properties
have a macroscopic (at least mesoscopic) nature of their origin, an
unambiguous scheme to describe the entire situation can only be found
when we can evaluate the lower limit at which these properties come
to being. Because it is in such pursuit that the evolution of each
of these properties over time and under a nonlinear context dependence
can be studied, which might ultimately provide us with the objective
information regarding who is contributing how much and due to precisely
how many number of atoms. The proposed scheme here achieves precisely
the same through the differential $eq^{n}-4$, by assigning residue-specific
atomic hydrophobicity values (or partial electrostatic charges, if
one attempts to study electrostatic contributions) to the variable
$\, x\,$, the contact-map information to the matrix $\, A\,$, and
known dependencies that describes effect of hydrophobicity (or electrostatics)
on binding free energy to function $\, f\left(x,t\right)\,$.\\
 \\
 \textbf{\underbar{2.3) Scope of applicability of the present
algorithm :}}\\
 Another example with polarizability studies might help in registering
the significance of present theory of biological observability. Poisson-Boltzmann
(PB) theory is a statistical mean field theory that characterizes
coarse-grained quantities such as the average particle distribution
function and the electrostatic potential together with thermodynamic
variables, in systems composed of many charged and point like particles
at thermal equilibrium. However, despite various modifications of
the scheme in dilute and strong coupling regimes {[}92, 93{]} and
numerous imaginative applications of it several paradigms of macromolecular
biological structures {[}94-96{]}; the statistical modeling of real
solutions - often in an intermediate regime - is still an open problem
{[}93,97{]}. It is an open problem because the precise threshold at
which emergence of the statistical properties take place is not clear
to us. Taking recourse to the present theory, we can attempt to achieve
clarity in our description of the aforementioned problem.\\
 \\
 \textbf{\underbar{2.4) Scope of applicability of the present
algorithm :}}\\
 Accurate measurement of residual dipolar couplings in weakly
aligned proteins can in principle provide incisive information about
the structure and dynamics of them in the solution state {[}98{]}.
But the problem in this operation stems from the nature of measured
information, which usually embody a convolution of the structural
and dynamic properties {[}99{]}. Amongst many other aspects, the sensitivity
of residual dipolar couplings to internal motions has been recognized
by many as an enormously interesting question {[}100-103{]}. This
is so because, unlike the conventional spin relaxation and chemical
exchange-based studies, residual dipolar couplings are sensitive to
motions spanning a wide range of time scales, and henceforth, they
might be considered as potent probes to monitor biologically relevant
motions{[}100{]}. But, many structure refinement protocols for analyzing
dipolar data implicitly assume that internal motions are either absent,
negligibly small, or uniform and axially symmetric in nature {[}104{]},{[}105{]}.
Although some sporadic attempts have been made to study dynamics in
protein interior, a rigorous theoretical framework that solves the
inverse problem \textbf{(P-3)}, namely, given the information regarding
residual dipolar coupling, to what extent can we observe the dynamics
of protein atoms, - has not been proposed. It isn't a question, because
these motions are not universally reflected in specific dipolar couplings
in presence of high anisotropy {[}99{]}, but a carefully designed
mathematical framework with top-down philosophy that can encompass
physical perturbations without delving into the bottom-up origins
of the later, can attempt to describe the situation with adequate
accuracy. The scope of this question can be meaningfully extended
if we notice that interference of existing internal motions with structural
interpretations of dipolar data - has not been studied with sufficient
thoroughness. Study with particular case {[}99{]} tends to suggest
that some internal motions can well be into the range of observability.
Here, from the context of the present algorithm, describing the change
in the magnitude of the (emergent) dipolar coupling as the dependent
variable of $eq^{n}-4$, the contact-map information for neighboring
atoms in the matrix $A$, and relevant information (either the residual
pKa values, or some cumulative measure of atomic partial charges at
residual level or any other parameter set that the user thinks pertinent)
in $\, f\left(x,t\right)\,$, - one can attempt to observe at which
threshold level of a number of atoms does the measured magnitude of
the residual dipolar coupling emerge.\\
 \\
 \\
 \textbf{\underbar{Applicability - 3) Case of NMR :}}\\
 Biomolecular structure determination through NMR starts with
collection of proton spectra. Protons in the various atoms in various
residues have different electronic environments {[}106{]}. The typical
electronic cloud dispositions give rise to typical local magnetic
fields which alter the static field $\, B_{0}\,$ to $\, B_{0}\left(1-\sigma\right)\,$,
altering therefore, the Larmor frequency of these protons. The resulting
proton spectrum is comprised of many peaks. These shifts in the Larmor
are characteristic of the chemical environment of the spins and are
termed chemical shifts (CS). Proteins are known to be flexible {[}3{]},
in solution, they undergo constant small conformational changes and
furthermore since the CSs are affected by tertiary structures {[}107-109{]},
we can regard the chemical shifts as dynamic (time-varying). However,
chemical shifts are typically viewed as a static property {[}110{]}
(largely due to the tools employed in traditional NMR analysis. The
NMR spectrometer records a series of time-domain signals, know as
Free Induction Decays (FIDs). A given atom's CS is encoded as a periodicity
within the FIDs. It is obtained by applying a Fourier Transform to
the FIDs. FIDs, being time-domain signals, are capable of encoding
CSs. However, it is not possible to observe CSs using the Fourier
Transform because the integration operation takes place over time).
Nevertheless, based on the CS information, it is possible to assign
the various peaks in the spectrum of a molecule to various protons,
a process called frequency labeling {[}111{]}. In theory, using these
shifts, the peaks in the spectrum can be uniquely assigned to protons
in various amino acids. Following this, a series of experiments are
used to selectively excite protons and study the effect on other protons.
This effect (Nuclear Overhauser effect) is proportional to $\, r^{\left(-6\right)}\,$
where r is the distance between nuclei. The obtained set of data then
provides the distance information between protons of various amino
acids. Once this distance is known, one can solve for a folded configuration
of the protein, which satisfied these distance constraints. The practical
scenario, on the other hand, presents a different picture since the
proton spectrum of a large protein molecule is obtained as one with
poor resolution {[}52{]} (because existence of large number of protons
leads to crowding of the spectra). Owing to this poor resolution,
the task of frequency labeling becomes extremely difficult.\\
 \\
 With the advent of 2D NMR {[}112{]} (here the connectivity between
distinct individual spins are delineated, and furthermore, the resonance
peaks are spread out in two dimensions leading to a substantial improvement
in peak separation, thus making the spectra far easier to interpret)
the aforementioned problem can be tackled. But the basic inverse problem
\textbf{(P-4)}, still lingers; viz., what should be the upper limit
of the number of residues (and atoms therein), so that given a proton
spectrum, the frequency labeling can be achieved ? Appropriate use
of the proposed scheme might help in more focused framing of \textbf{(P-4)},
before an attempt to obtain the solution for it.\\
 However, we must mention here that control theoretic applications
to NMR studies are not new {[}52{]} and many a successful constructs
have been proposed banking on the pure mathematical studies with basic
knowledge of Physics that involves little or no computational prowess.
However, in terms of the scope and orientation, the approach proposed
in this work is first of its kind. \\
 \\
 \\
 \textbf{\underbar{Applicability - 4) Case of Drug-Discovery and
Computational-Chemistry :}}\\
 In the paradigm of drug discovery (and computational chemistry,
in general) an outstanding problem can be stated as, given the information
regarding the structure of a protein active site and a list of potential
small molecule ligands, predict the binding mode and estimate the
binding affinity for each ligand {[}113{]}. This problem has multiple
aspects associated with it and has been a field of intense computational
and experimental analysis over the last fifteen years. However, certain
basic questions in this paradigm still remain unanswered and in the
absence of theoretically deduced unambiguous criteria set, approaches
to these questions often provide inconsistent results {[}114{]}. The
entire operation of docking can be summarized into two operations;
first, the operation of {}``posing''; viz., the appropriate positioning
of the correct conformer of a ligand in the active site (combination
of conformation and orientation being known as a {}``pose''). Second,
the operation of {}``scoring'', where poses are selected and ranked
with respect to some scoring function {[}113-114{]}. Although apparently
straightforward, this two step process involves many a complex (non-linear)
physico-chemical interactions from geometric perspective and in their
bid to simultaneously address these issues through this two-step process,
many approximations and inadequate constructs are resorted to. These
have been identified in many recent works {[}114-117{]}. (For example,
simplistic treatment of electrostatics, electronic polarization, aqueous
desolvation, and ionic influences; lack of accounting for entropy
changes in the protein and the ligand on binding; inadequate weighting
of proton positions (tautomers, rotamers) and charge states (ionization)
of both protein and ligand; assumptions in many (but not all) of the
cases that active site is rigid (possibly including tightly bound
water molecules) and that only the small molecule can move; etc ..).
A close scrutiny amongst these drawbacks points immediately to an
underlying connecting factor. Many of these shortcomings exist because
the precise mode of emergence of these features from a certain set
of number of atoms, is not known; furthermore, the (non-linear) dependencies
that these properties might be having on one another is difficult
to decipher too, because of the same reason. A solution to these problems
can be found from examining the situation from a coherent perspective
where the distinguishability of a non-statistical (non-macroscopic,
non-thermodynamic) system of atoms can be ensured; but at the same
time, conditions for observability of the emergence of macroscopic
(statistical) properties are appropriately identified. The control
theoretic approach presented in this work, might help in quantifying
many of these features, from the perspective of inverse problem; where
the atomic origin of these features will be addressed from a bottom-up
mathematical standpoint without delving into the depths of physical
and/or chemical dependencies. For example, a computational framework
can be set-up where the user assigns measured (computational or experimental)
change of entropy of either protein or ligand to the variable $\dot{\, x}\,$,
the neighboring atom information of it in the form of a contact-map
to the matrix $A$, the conformational fluctuation based information
for each amino acids involved, to $\, x\,$; and finally, some relevant
parameter that maps residual conformational fluctuations with (macroscopic)
entropy, to the nonlinear context-dependent function $\, f\left(x,t\right)\,$.
(Otherwise, the cumulative effect of conformational fluctuations,
as the internal energy of the set of atoms under consideration, can
be mapped to entropy and assigned to $\, f\left(x,t\right)\,$.) With
such a scheme, for a known set of $\dot{\, x}\,$ values, the corresponding
$\, x\,$; or the more direct; known $\, x\,$ to unknown $\dot{\, x}\,$
studies - can be attempted. - Obviously, an expert in this sphere
of knowledge can ascertain the feasibility of attempting certain problems,
the general mathematical framework stays valid for every set of parameter.
\\
 \\
 On the other hand, the scoring functions serve as objective function
to classify diverse poses of a single ligand in the receptor binding
site before estimating the binding affinities of different receptor\textendash{}ligand
complexes (and ranking them) upon the docking of a compound database
is performed {[}118{]}. Interestingly, the drawbacks of scoring functions,
as elaborated in a recent work {[}119{]} (failure to accommodate subtle
physical effects affecting the experimental binding energy; viz.,
the treatment of polar groups in the ligand or the protein being desolvated
upon binding but failing to find a matching polar interaction in the
complex, treatment of hydrophobic patches of the ligand exposed to
the solvent upon binding, a more comprehensive treatment of loss of
rotational and translational entropy; etc ..) - also suffer from the
same nature of problems as the ones explained in that last paragraph.
Here also, a consistent scheme that does not involve itself with the
mind-boggling complexity of the physico-chemical interactions, but
circumvents it by assuming that all the aforementioned properties
come to being due to some or the other form of interatomic interactions
between a set of atoms involving electromagnetic forces, before attempting
to identify the number of atoms necessary to produce the property
under consideration - can be of extreme utility. Since the algorithm
proposed here targets the transition zone between nano-scale individualistic
properties to mesoscopic and subsequently macroscopic properties,
by targeting the number of interacting atoms rather than the property
itself; - it can overlook the complex physico-chemical details. Yet,
it can monitor, from which threshold of atoms, the emergence of a
particular property is observed. This helps him to identify the possible
dependencies one property can have on the others and predict which
ones are more fundamental than the others. \\

\section{\underbar{Conclusion :}}

An algorithm to study the lower threshold of emergence for various
biophysical properties, is presented in this work. Categorical linkages
between rigorous mathematical backbone with protein biophysical properties
are established. An exact knowledge of these limits hold paramount
utiliterian importance in the paradigm of the nascent field Nano-Bioscience.
They, on the other hand, provide the contemporary state of protein
interior knowledge with constructs to investigate the fundamental
questions of protein biophysics. In near future, when the computational
facilities become less prohibitive, these algorithms can be implemented
to answer the questions like {}``precisely how many atoms are necessary
for us to observe hydrophobicity in a protein under a specified biological
context''? \\
\\
\\
\textbf{\underbar{Acknowledgment :}} This study was supported
by COE scheme, Department of Biotechnology, Government of India.\\
 One of the authors, Anirban, wants to thank Om Prakash Pandey,
for all his efforts to make the author understand the key concepts
of drug-discovery, as have been discussed in this work.\\
 \\
 \\
 \textbf{\underbar{References :}}\\
 {[}01{]} Vendruscolo M, Dokholyan NV, Paci E, Karplus M (2002)
Small-world view of the amino acids that play a key role in protein
folding. Phys Rev E 65 : 061910.\\
 {[}02{]} Kaya H, Chan HS (2000) Energetic Components of Cooperative
Protein Folding. Phys Rev Lett : 85 : 4823-4826.\\
 {[}03{]} Fitzkee NC, Fleming PJ, Gong H, Panasik N Jr, Street
TO, Rose GD (2005) Are proteins made from a limited parts list? TRENDS
in Biochemical Sciences 30 : 73-80.\\
 {[}04{]} Tissen J, Fraaije J, Drenth J, Berendsen H (1994) Mesoscopic
theories for protein crystal growth. Acta Cryst. D 50 : 569-571.\\
 {[}05{]} Phillips JC (2009) Scaling and self-organized criticality
in proteins I. Proc Natl Acad Sci USA. 106 : 3107-3112.\\
 {[}06{]} Phillips JC (2009) Scaling and self-organized criticality
in proteins II; Proc Natl Acad Sci USA 106 : 3113\textendash{}3118.\\
 {[}07{]} Banerji A, Ghosh I (2009) A new computational model
to study mass inhomogeneity and hydrophobicity inhomogeneity in proteins;
Eur. Biophys J epub prior publication; DOI 10.1007/s00249-009-0409-1.\\
 {[}08{]} Rother K, Preissner R, Goede A, Frömmel C (2003) Inhomogeneous
molecular density: reference packing densities and distribution of
cavities within proteins. Bioinformatics 19 : 2112-2121.\\
 {[}09{]} Reuveni S, Granek R, Klafter J (2008) Proteins: Coexistence
of Stability and Flexibility; Phys Rev Lett 100 : 208101.\\
 {[}10{]} MacDonald M, Jan N (1986) Fractons and the fractal dimension
of proteins. Canadian J Phys 64 : 1353-1355.\\
 {[}11{]} Li H., Chen S, Zhao H (1998) Fractal mechanisms for
the allosteric effects of proteins and enzymes; Biophys J 58 : 1313-1320.\\
 {[}12{]} Lewis M, Rees DC (1985) Fractal Surfaces of Proteins.
Science 230 : 1163-1165.\\
 {[}13{]} Millhauser GL, Salpeter EE, Oswald RE (1988) Diffusion
models of ion-channel gating and the origin of the power-law distributions
from single-channel recordings. Proc. Natl. Acad Sci. USA. 85 : 1503-1507.\\
 {[}14{]} Helman JS, Coniglio A, Tsallis C (1984) Fractals and
the fractal structure of proteins. Phys Rev Lett 53 : 1195-1197.\\
 {[}15{]} Dewey TG (1994) Fractal analysis of proton exchange
kinetics in lysozyme (protein dynamics/fractal reaction kinetics)
Proc Natl Acad Sci USA 91 : 12101-12104.\\
 {[}16{]} Dewey TG (1999) Statistical Mechanics of Protein Sequences.
Phys Rev E 60 : 4652-4658.\\
 {[}17{]} Medini D, Widom A (2003) Atomic scale fractal dimensionality
in proteins. J Chem Phys 118 : 2405-2410.\\
 {[}18{]} Enright MB, Leitner DM (2005) Mass fractal dimension
and the compactness of proteins. Phys Rev E 71 : 011912.\\
 {[}19{]} Banerji A, Ghosh I (2009) Revisiting the Myths of Protein
Interior; Submitted Manuscript.\\
 {[}20{]} Xiong Liu and Hassan A. Karimi (2007) High-throughput
modeling and analysis of protein structural dynamics; Briefings in
Bioinformatics; 8(6):432-445. \\
 {[}21{]} K. A. Peterson, M. B. Zimmt, S. Linse, R. P. Domingue,
and M. D. Fayer (1987) Quantitative Determination of the Radius of
Gyration of Poly(methy1 methacrylate) in the Amorphous Solid State
by Time-Resolved Fluorescence Depolarization Measurements of Excitation
Transport; Macromolecules; 20; 168-175.\\
 {[}22{]} Annunziata O., Buzatu D., and Albright J.G. (2005) Protein
Diffusion Coefficients Determined by Macroscopic-Gradient Rayleigh
Interferometry and Dynamic Light Scattering; Langmuir; 21 (26); 12085\textendash{}12089.\\
 {[}23{]} Themis Lazaridis, Martin Karplus (2003) Thermodynamics
of protein folding: a microscopic view; Biophysical Chemistry; 100;
367\textendash{}395.\\
 {[}24{]} William Bialek and Rama Ranganathan (2007) Rediscovering
the power of pairwise interactions; arXiv:0712.4397v1; 2007.\\
 {[}25{]} M Socolich, SW Lockless, WP Russ, H Lee, KH Gardner
and R Ranganathan (2005) Evolutionary information for specifying a
protein fold. Nature 437, 512\textendash{} 518.\\
 {[}26{]} Amadei, A., Linssen, A.B.M. and Berendsen, H.J.C. (1993)
Essential dynamics of proteins. Proteins, 17, 412\textendash{}425.\\
 {[}27{]} Teeter,M.M. and Case,D.A. (1990) Harmonic and quasiharmonic
descriptions of crambin. J. Phys. Chem., 94, 8091\textendash{}8097.
\\
{[}28{]} Brooks,B. and Karplus,M. (1983) Harmonic dynamics of
proteins: normal modes and fluctuations in bovine pancreatic trypsin
inhibitor. Proc. Natl Acad. Sci. USA, 80, 6571\textendash{}6575.\\
 {[}29{]} Go,N., Noguti,T. and Nishikawa,T. (1983) Dynamics of
a small globular protein in terms of low-frequency vibrational modes.
Proc. Natl Acad. Sci. USA, 80, 3696\textendash{}3700.\\
 {[}30{]} Levitt, M., Sander C. and Stern P. S (1985) Protein
Normal-Mode Dynamics: Trypsin Inhibitor, Crambin, Ribonuclease and
Lysozyme. J. Mol. Biol. 181, 423-447.\\
 {[}31{]} Yang, H., Luo, G., Karnchanaphanurach, P., Louie, T.
M., Rech, I., Cova, S., Xun, L. and Xie, X. S. (2003) Science 302,
262\textendash{}266.\\
 {[}32{]} Steven T. Whitten, Bertrand Garc\i{}a-Moreno E, and
Vincent J. Hilser (2005) Local conformational fluctuations can modulate
the coupling between proton binding and global structural transitions
in proteins; PNAS; 102(12); 4282\textendash{} 4287.\\
 {[}33{]} Van den Berg, B., Ellis, R.J., and Dobson, C.M. (1999)
Effects of macromolecular crowding on protein folding and aggregation.
EMBO J. 18: 6927\textendash{}6933.\\
 {[}34{]} Kim, P.S. and Baldwin, R.L. (1982) Specific intermediates
in the folding reactions of small proteins and the mechanism of protein
folding. Annu. Rev. Biochem. 51: 459\textendash{}489.\\
 {[}35{]} Kim, P.S. and Baldwin, R.L. (1990) Intermediates in
the folding reactions of small proteins. Annu. Rev. Biochem. 59: 631\textendash{}660.\\
 {[}36{]} P.F.W. Stouten, C. Frommel, H. Nakamura, C. Sander (1993)
An effective solvation term based on atomic occupancies for use in
protein simulations, Mol. Simul. 10 (1993) 97\textendash{}120.\\
 {[}37{]} S. Miller, J. Janin, A.M. Lesk, C. Chothia (1987) Interior
and surface of monomeric proteins, J. Mol. Biol. 196: 641\textendash{}656.\\
 {[}38{]} G.I. Makhatadze, P.L. Privalov (1995) Energetics of
protein structure, Adv. Protein Chem. 47 (1995) 307\textendash{}425.\\
 {[}39{]} T. Lazaridis, G. Archontis, M. Karplus (1995) Enthalpic
contribution to protein stability: atom-based calculations and statistical
mechanics, Adv. Protein Chem. 47: 231\textendash{}306, (Academic Press,
Inc.).\\
 {[}40{]} K.A. Dill, D. Shortle (1991) Denatured states of proteins,
Ann. Rev. Biochem. 60 : 795.\\
 {[}41{]} D. Shortle, The denatured state (the other half of the
folding equation) and its role in protein stability, FASEB J. 10 (1996)
27\textendash{}34.\\
 {[}42{]} E.D. Sontag. Some new directions in control theory inspired
by systems biology. IEEE Proc. Systems Biology, 1:9\textendash{}18,
2004.\\
 {[}43{]} E.D. Sontag. Molecular systems biology and control.
Eur. J. Control, 11(4-5):396\textendash{}435, 2005.\\
 {[}44{]} E.D. Sontag. Some Remarks on Input Choices for Biochemical
Systems. arXiv:math/0606261v1 ; 2006.\\
 {[}45{]} L.A. Torres, V. Ibarra-Junquera, P. Escalante-Minakata
and H.C. Rosu; High-Gain Nonlinear Observer for Simple Genetic Regulation
Process; Physica A 380 (1 July 2007) 235-240.\\
 {[}46{]} J.P. Gauthier and I.A.K. Kupka, Observability and observers
for nonlinear systems, SIAM J. Control Optimization 32 (1994) 975-994.\\
 {[}47{]} A.J. Krener and A. Isidori, Linearization by output
injection and nonlinear observers, Systems \& Control Letters, 3 (1983)
47-52.\\
 {[}48{]} J.L. Willems, Stability Theory of Dynamical Systems,
Thomas Nelson and Sons Ltd., London, 1970; pp 52-53.\\
 {[}49{]} M. Green and D.J.N. Limebeer, Linear Robust Control,
Prentice Hall, 1995. pp-82. \\
 {[}50{]} S. Camalet, T. Duke, F. Julicher, and J. Prost. Auditory
sensitivity provided by self-tuned critical oscillations of hair cells.
Proc. of Natl. Acad. Sc. U.S.A., 97(7):3183\textendash{}3188, 2000.\\
 {[}51{]} E.D. Sontag. Some new directions in control theory inspired
by systems biology. Systems Biology, 1(1):9\textendash{}18, 2004.\\
 {[}52{]} Mabuchi H. and Khaneja N.; Principles and applications
of control in quantum systems; Int. J. Robust Nonlinear Control 2004;
15(15); 647-667.\\
 {[}53{]} I. Miroshnik, V. Nikiforov, and A. Fradkov. Nonlinear
and Adaptive Control of Complex Systems. Kluwer, 1999.\\
 {[}54{]} T.T. Georgiou and M.C. Smith; Feedback Control and the
Arrow of Time; arXiv:0804.3117v1 ; 2008. \\
 {[}55{]} Graziano G.; Hydrophobicity of benzene; Biophysical
Chemistry; 1999; 82; 69-79.\\
 {[}56{]} Ashbaugh H.S., Asthagiri D., Pratt L.R., and Rempe S.B.;
Hydration of Krypton and Consideration of Clathrate Models of Hydrophobic
Effects from the Perspective of Quasi-Chemical Theory; Biophys Chem.
2003; 105(2-3): 323-338.\\
 {[}57{]} Eisenberg D, Weiss RM, Terwilliger TC.; The hydrophobic
moment detects periodicity in protein hydrophobicity.; Proc Natl Acad
Sci U S A. 1984; 81(1):140-144.\\
 {[}58{]} Southall, N. T.; Dill, K. A. The mechanism of hydrophobic
solvation depends on solute radius. J. Phys. Chem. B 2000, 104, 1326-1331.\\
 {[}59{]} Cheng, Y.-K.; Rossky, P. J. Surface topography dependence
of biomolecular hydrophobic hydration. Nature 1998, 392, 696-699.\\
 {[}60{]} THEMIS LAZARIDIS; Solvent Size vs Cohesive Energy as
the Origin of Hydrophobicity; ACCOUNTS OF CHEMICAL RESEARCH; 2001;
34(12)\\
 {[}61{]} Seishi Shimizu and Hue Sun Chan; Temperature dependence
of hydrophobic interactions: A mean force perspective, effects of
water density, and nonadditivity of thermodynamic signatures; JOURNAL
OF CHEMICAL PHYSICS; 2000; 113(11); 4683-4700. \\
 {[}62{]} Seishi Shimizu and Hue Sun Chan; Temperature dependence
of hydrophobic interactions: A mean force perspective, effects of
water density, and nonadditivity of thermodynamic signatures; JOURNAL
OF CHEMICAL PHYSICS; 2000; 113(11); 4683-4700.\\
 {[}63{]} R.H. Wood and P.T. Thompson; Differences between pair
and bulk hydrophobic interactions; Proc. Natl. Acad. Sci. USA; 1990;
87; 946-949.\\
 {[}64{]} A. Sacco and M. Holz; NMR Studies on Hydrophobic Interactions
in Solution. Part 2: Temperature and Urea Effect on the Self-association
of Ethanol in Water J. Chem. Soc., Faraday Trans.; 93; 1101-1104;
1997.\\
 {[}65{]} W. Blokzijl and J.B.F.N. Engberts, Hydrophobic Effects.
Opinions and Facts; Angewandte Chemie (international edn in English);
32 (11), 1545-1579.\\
 {[}66{]} S. Ludemann, R. Abseher, H. Schreiber, and O. Steinhauser;
The Temperature-Dependence of Hydrophobic Association in Water. Pair
versus Bulk Hydrophobic Interactions; J. Am. Chem. Soc. 119; 4206-4213;
1997.\\
 {[}67{]} D. van Belle and S.J. Wodak; Molecular dynamics study
of methane hydration and methane association in a polarizable water
phase; J. Am. Chem. Soc. 115; 647-652; 1993.\\
 {[}68{]} G. Hummer, S. Garde, A.E. Garcia, M.E. Paulatis, and
L.R. Pratt; The pressure dependence of hydrophobic interactions is
consistent with the observed pressure denaturation of proteins; Proc.
Natl. Acad. Sci. USA 95, 1552-1555; 1998.\\
 {[}69{]} Vladimir P. Denisov, Jamie L. Schlessman, Bertrand Garc\i{}a-Moreno
E., and Bertil Halle; Stabilization of Internal Charges in a Protein:
Water Penetration or Conformational Change? Biophysical Journal; 2004;
87; 3982\textendash{}3994.\\
 {[}70{]} Warshel, A. 1981. Calculations of enzymatic reactions:
calculations of pKa, proton transfer reactions, and general acid catalysis
reactions in enzymes. Biochemistry. 20:3167\textendash{}3177.\\
 {[}71{]} Antosiewicz, J., J. A. McCammon, and M. K. Gilson. 1996.
The determinants of pKas in proteins. Biochemistry. 35:7819\textendash{}7833.\\
 {[}72{]} Schutz, C. N., and A. Warshel. 2001. What are the dielectric
\textquotedbl{}constants\textquotedbl{} of proteins and how to validate
electrostatic models? Proteins. 44: 400\textendash{}417.\\
 {[}73{]} Forsyth, W. R., J. M. Antosiewicz, and A. D. Robertson.
2002. Empirical relationships between protein structure and carboxyl
pKa values in proteins. Proteins. 48:388\textendash{}403.\\
 {[}74{]} Simonson, T., J. Carlsson, and D. A. Case. 2004. Proton
binding to proteins: pKa calculations with explicit and implicit solvent
models. J. Am. Chem. Soc. 126 : 4167\textendash{}4180.\\
 {[}75{]}Song, X.J.; An inhomogeneous model of protein dielectric properties
: Intrinsic polarizabilities of amino acids. Chem. Phys. 116, 2 (2002).\\
 {[}76{]} Talley K., Ng C., Shoppell M., Kundrotas P., and Alexov
E.; On the electrostatic component of protein-protein binding free
energy; PMC Biophysics 2008, 1:2.\\
 {[}77{]} Simonson, T., and C. L. Brooks. 1996. Charge screening
and the dielectric constant of proteins: insights from molecular dynamics.
J. Am. Chem. Soc. 118:8452\textendash{}8458.\\
 {[}78{]} Pitera, J. W., M. Falta, and W. F. van Gunsteren. 2001.
Dielectric properties of proteins from simulation: the effects of
solvent, ligands, pH, and temperature. Biophys. J. 80:2546\textendash{}2555.\\
 {[}79{]} Honig, B., and A. Nicholls. 1995. Classical electrostatics
in biology and chemistry. Science. 268:1144\textendash{}1149.\\
 {[}80{]} Bashford, D., and D. A. Case. 2000. Generalized Born
models of macromolecular solvation effects. Annu. Rev. Phys. Chem.
51:129\textendash{}152.\\
 {[}81{]} Jin Aun Ng, Taira Vora, Vikram Krishnamurthy and Shin-Ho
Chung; Estimating the dielectric constant of the channel protein and
pore; European Biophysics Journal;37(2), 2008.\\
 {[}82{]} Jensen JH: Calculating pH and salt dependence of protein-protein
binding. Curr Pharm Biotechnol 2008, 9(2):96-102.\\
 {[}83{]} Bertonati C, Honig B, Alexov E: Poisson-Boltzmann calculations
of nonspecific salt effects on protein-protein binding free energies.
Biophys J 2007, 92(6):1891-1899.\\
 {[}84{]} Talley K, Alexov E: Modelling Salt Dependence of Protein-Protein
Association:Linear vs Non-Linear Poisson-Bolzmann Equation. Comm in
Computational Physics 2008, 4:169-179.\\
 {[}85{]} Bordner AJ, Abagyan R: Statistical analysis and prediction
of protein-protein interfaces. Proteins 2005, 60(3):353-366.\\
 {[}86{]} Muegge I, Schweins T, Warshel A: Electrostatic Contribution
to Protein-Protein Binding Affinities: Application to Rap/Raf Interaction.
Proteins 1998, 30:407-423.\\
 {[}87{]} Lee LP, Tidor B: Barstar is electrostatically optimized
for tight binding to barnase. Nat Struct Biol 2001, 8(1):73-76.\\
 {[}88{]} Sims PA, Wong CF, McCammon JA: Charge optimization of
the interface between protein kinases and their ligands. J Comput
Chem 2004, 25(11):1416-1429.\\
 {[}89{]} Norel R, Sheinerman F, Petrey D, Honig B: Electrostatic
contribution to protein-protein interactions: Fast energetic filters
for docking and their physical basis. Prot Sci 2001, 10:2147-2161.\\
 {[}90{]} Dong F, Vijayakumar M, Zhou HX: Comparison of calculation
and experiment implicates significant electrostatic contributions
to the binding stability of barnase and barstar. Biophys J 2003, 85(1):49-60.\\
 {[}91{]} Sheinerman FB, Honig B: On the role of electrostatic
interactions in the design of protein-protein interfaces. J Mol Biol;
2002, 318(1):161-177.\\
 {[}92{]} Borukhov I., Andelman D., and Orland H.; Steric Effects
in Electrolytes: A modified Poisson-Boltzmann equation; Phys. Rev.
Lett.; 1997; 79; 435.\\
 {[}93{]} Fabien Paillusson, Maria Barbi, Jean-Marc Victor; Poisson-Boltzmann
for oppositely charged bodies: an explicit derivation; arXiv:0902.1457v1
{[}cond-mat.soft{]} 2009.\\
 {[}94{]} Tang, C., Iwahara, J. and Clore, G. M. Visualization
of transient encounter complexes in protein protein association; Nature;
2006; 444 : 383-386.\\
 {[}95{]} von Hippel, H. Diffusion driven mechanism of protein
translocation on nucleic acids. III. The E. coli lac repressor-operator
interaction: kinetic measurements and conclusions; Biochemistry; 2007;
20; 6961-6977.\\
 {[}96{]} Sens P. and Joanny J.-F. Counterion Release and Electrostatic
Adsorption Phys. Rev. Lett. 84 4862-4865 (2000). \\
 {[}97{]} Netz R. R, and Orland H.; Beyond Poisson Boltzmann:
Fluctuations effects and correlation functions. Eur. Phys. J. E; 2000;
1; 203-214.\\
 {[}98{]} Tolman, J.R., Flanagan, J.M., Kennedy, M.A., Prestegard,
J.H; Nuclear magnetic dipole interactions in field-oriented proteins:
information for structure determination in solution; Proc. Natl. Acad.
Sci. U.S.A. 1995, 92, 9279-9283.\\
 {[}99{]} Joel R. Tolman, Hashim M. Al-Hashimi, Lewis E. Kay,
and James H. Prestegard; Structural and Dynamic Analysis of Residual
Dipolar Coupling Data for Proteins; J. Am. Chem. Soc. 2001, 123, 1416-1424.\\
 {[}100{]} Tolman JR, Flanagan JM, Kennedy MA, Prestegard JH;
NMR evidence for slow collective motions in cyanometmyoglobin; Nat
Struct Biol. 1997 Apr;4(4):292-297.\\
 {[}101{]} Kay, L.E.; Protein dynamics from NMR; Nat. Struct.
Biol. 1998, 5, 513-517.\\
 {[}102{]} Foster MP, McElroy CA, Amero CD; Solution NMR of large
molecules and assemblies; Biochemistry; 2007; 46(2); 331-340.\\
 {[}103{]} Vogeli B, Yao L.; Correlated dynamics between protein
HN and HC bonds observed by NMR cross relaxation; J Am Chem Soc. 2009;
131(10); 3668-3678.\\
 {[}104{]} Clore GM, Gronenborn AM; New methods of structure refinement
for macromolecular structure determination by NMR; Proc Natl Acad
Sci U S A.; 1998; 95(11); 5891-5898.\\
 {[}105{]} Clore, G.M. \& Garrett, D.S. (1999) R-factor, Free
R and complete cross-validation for dipolar coupling refinement of
NMR structures. J. Am. Chem. Soc. 121, 9008-9012. \\
 {[}106{]} Peter Edwards, Derek Sleeman, Gordon C.K. Roberts \&
Lu Yun Lian; {}``An AI Approach to the Interpretation of the NMR
Spectra of Proteins''; Artificial intelligence and molecular biology;
1993; American Association for Artificial Intelligence Menlo Park,
CA, USA; pp 396-432. \\
 {[}107{]} Sitkoff, D., \& Case, D. Density Functional Calculations
of Proton Chemical Shifts in Model Peptides. J. Am. Chem Soc. 119,
12262-12273 (1997).\\
 {[}108{]} Dejaegere, A. P., Case, D. Density Functional Study
of Ribose and Deoxyribose Chemical Shifts, J. Phys. Chem. A 102, 5280-5289
(1998).\\
 {[}109{]} Moravetski, V., Hill, J. R., Eichler, U., Sauer, J.
29 Si NMR Chemical Shifts of Silicate Species: Ab Initio Study of
Environment and Structure Effects, J. Am. Chem. Soc. 118, 13015-13020
(1996).\\
 {[}110{]} C. J. Langmead and B. R. Donald; Extracting Structural
Information Using Time-Frequency Analysis of Protein NMR Data; Proceedings
of The Fifth Annual International Conference on Computational Molecular
Biology (RECOMB), Montreal, April 22-25 (2001) pp. 164-175.\\
 {[}111{]} K. Wuthrich, NMR of Proteins and Nucleic Acids, Wiley
\& Sons (1986).\\
 {[}112{]} R.M. Cooke \& I.D. Campbell, Protein Structure Determination
by Nuclear Magnetic Resonance, BioEssays, 1988, 8 (2), 52-56.\\
 {[}113{]} Vigers G.P.A., and Rizzi J.P.; Multiple Active Site
Corrections for Docking and Virtual Screening; J. Med. Chem.; 2004;
47; 80-89.\\
 {[}114{]} Hawkins P.C.D., Skillman A.G., and Nicholls A.; Comparison
of Shape-Matching and Docking as Virtual Screening Tools; J. Med.
Chem.; 2007; 50; 74-82.\\
 {[}115{]} Warren, G.L.; et al. A critical assessment of docking
programs and scoring functions. J. Med. Chem. 2006, 49, 5912-5931.\\
 {[}116{]} Maiorov, V; Sheridan, R.P.; Enhanced virtual screening
by combined use of two docking methods: Getting the most on a limited
budget. J. Chem. Inf. Model. 2005, 45, 1017-1024.\\
 {[}117{]} Marsden, P.M.; Puvenendrampillai, D.; Mitchell, J.B.O.;
Predicting protein-ligand binding affinities: a low scoring game?
Org. Biomol. Chem. 2004, 2, 231-237.\\
 {[}118{]} Schulz-Gasch T., Stahl M.; Binding site characteristics
in structure-based virtual screening: evaluation of current docking
tools; J Mol Model (2003) 9:47\textendash{}57.\\
 {[}119{]} Jacobsson M., and Karlen A.; Ligand Bias of Scoring
Functions in Structure-Based Virtual Screening; J. Chem. Inf. Model.;
2006; 46; 1334-1343. 
\end{document}